\documentclass[%
superscriptaddress,
reprint,
showpacs,
 amsmath,amssymb,
 aps,
prl,
]{revtex4-1}

\newcommand\beq{\begin{equation}}
\newcommand\eeq{\end{equation}}

\usepackage{graphicx}
\usepackage{dcolumn}
\usepackage{bm}

\begin{document}


\title{Nonlocal Transformation Optics}

\author{Giuseppe Castaldi}
\author{Vincenzo Galdi}
\email{vgaldi@unisannio.it}
\affiliation{Waves Group, Department of Engineering, University of Sannio, I-82100 Benevento, Italy
}%

\author{Andrea Al\`u}
\affiliation{Department of Electrical and Computer Engineering, The University of Texas at Austin, Austin, TX 78712, USA}%

\author{Nader Engheta}
\affiliation{Department of Electrical and Systems Engineering, University of Pennsylvania, Philadelphia, PA 19104, USA}%

\date{\today}


\begin{abstract}
We show that the powerful framework of transformation optics may be exploited for engineering the nonlocal response of artificial electromagnetic materials. Relying on the form-invariant properties of coordinate-transformed Maxwell's equations in the spectral domain, we derive the general constitutive ``blueprints'' of transformation media yielding prescribed nonlocal field-manipulation effects, and provide a physically-incisive and powerful geometrical interpretation in terms of deformation of the equi-frequency contours. In order to illustrate the potentials of our approach, we present an example of application to a wave-splitting refraction scenario, which may be implemented via a simple class of artificial materials. Our results provide a systematic and versatile framework which may open intriguing venues in dispersion engineering of artificial materials.
\end{abstract}

\maketitle

Spatial dispersion, i.e., the {\em nonlocal} character of the electromagnetic (EM) constitutive relationships \cite{Landau,Halevi}, is typically regarded as a negligible effect for most natural media. However, there is currently a growing interest in its study, in view of its critical relevance in the homogenized (effective-medium) modeling of many artificial EM materials of practical interest \cite{Alu0} (based, e.g., on small resonant scatterers \cite{Belov1,Silveirinha}, wires \cite{Simovski1,Belov2,Pollard}, layered metallo-dielectric composites \cite{Elser,Belov3}, etc.), as well as in a variety of related effects including artificial magnetism \cite{Alu1}, wave splitting into multiple beams \cite{Belov3,Kozyrev,Shen1}, beam tailoring \cite{Shen2}, and ultrafast nonlinear optical response \cite{Wurtz}. If, for most metamaterials, spatial dispersion is seen as a nuisance, counterproductive for practical applications \cite{Luukkonen}, its proper tailoring and engineering may add novel degrees of freedom in the wave interaction with complex materials \cite{Notomi}.

The transformation optics (TO) paradigm \cite{Leonhardt,Pendry} has rapidly established itself as a very powerful and versatile approach to the systematic design of artificial materials with assigned field-manipulation capabilities (see, e.g., \cite{Chen,Shalaev} for recent reviews). Standard TO basically relies on the form-invariant properties of coordinate-transformed Helmholtz \cite{Leonhardt} and Maxwell's equations \cite{Pendry}. Recently, alternative approaches have been proposed, based, e.g., on direct field transformations \cite{Tretyakov}, triple-space-time transformations \cite{Bergamin}, and complex coordinate mapping \cite{Castaldi}, which generalize and extend the class of ``transformation media'' (e.g., to nonreciprocal, bianisotropic, single-negative, indefinite, and moving media) that may be obtained. Although the approach in \cite{Bergamin} seems potentially capable to account for nonlocal effects, attention and applications have been hitherto focused on local transformation media. 
In this paper, we propose to apply the TO approach to develop a systematic framework for the engineering of nonlocal artificial
materials, paving the way to novel metamaterial devices and applications.

Our proposed approach is based on coordinate transformations in the spectral (wavenumber) domain, where nonlocal constitutive relationships are most easily dealt with in terms of wavenumber-dependent constitutive operators. For simplicity, we start considering a distribution of electric and magnetic sources (${\bf J}'$ and ${\bf M'}$, respectively) radiating an electromagnetic (EM) field (${\bf E}'$, ${\bf H'}$) in a vacuum auxiliary space, identified by primed coordinates ${\bf r}'\equiv(x',y',z')$. In the time-harmonic [$\exp(-i\omega t)$] regime, and introducing the spatial Fourier transform 
\beq
{\bf {\tilde G}}'({\bf k}')= \int {\bf G}'({\bf r}') \exp \left(
-i{\bf k}' \cdot {\bf r}'  
\right) d{\bf r}'
\label{eq:FT},
\eeq
the relevant Maxwell's curl equations can be fully algebrized in the spectral (${\bf k}'$) domain, viz., 
\begin{subequations}
\begin{eqnarray}
i{\bf k}'\times {\bf {\tilde E}}' ({\bf k}') = i\omega \mu_0 {\bf {\tilde H}}'({\bf k}') - {\bf {\tilde M}}'({\bf k}'),\\
i{\bf k}'\times {\bf {\tilde H}}'({\bf k}')  = -i\omega \varepsilon_0 {\bf {\tilde E}}'({\bf k}') + {\bf {\tilde J}}'({\bf k}'),  
\end{eqnarray}
\label{eq:Maxwell}
\end{subequations}
with $\epsilon_0$ and $\mu_0$ denoting the vacuum electrical permittivity and magnetic permeability, respectively.
Throughout the paper, boldface symbols identify vector quantities, and the tilde $\sim$ identifies spectral-domain quantities.

We now introduce a real-valued coordinate transformation to a new spectral domain ${\bf k}$,
\beq
{\bf k}'= {\tilde {\underline {\underline \Lambda}}}^T({\bf k})  \cdot {\bf k}={\tilde {\bf F}}\left({\bf k}\right),
\label{eq:mapping}
\eeq
with the double underline identifying a second-rank tensor operator, and the superscript $^T$ denoting the transpose. Similar to the spatial-domain TO, we exploit the form-invariant properties of Maxwell's equations in the mapped spectral domain ${\bf k}$ [and associated, via (\ref{eq:FT}), spatial domain ${\bf r}\equiv(x,y,z)$] in order to relate the corresponding fields (${\bf E}$, ${\bf H}$), sources (${\bf J}$, ${\bf M}$), and constitutive relationships (in terms of relative permittivity and permeability tensors ${\tilde  {\underline {\underline \varepsilon}}}$ and ${\tilde  {\underline {\underline \mu}}}$, respectively) to those in vacuum [cf. (\ref{eq:Maxwell})]
\begin{subequations}
\begin{eqnarray}
\left\{
{\tilde {\bf E}},{\tilde {\bf H}}
\right\}
({\bf k}) &=& {\tilde {\underline {\underline \Lambda}}}^{-T}({\bf k})
  \cdot \left\{
{\tilde {\bf E}}',{\tilde {\bf H}}'
\right\}
  \left[
{\tilde {\bf F}}\left({\bf k}\right)
 \! \right],\label{eq:field}\\
\!\!\left\{
{\tilde {\bf J}},{\tilde {\bf M}}
\right\}\!
({\bf k}) &\!\!=\!& 
\!\!\!\!\!\det^{~~~~~-1}\!\left[{\tilde {\underline {\underline \Lambda}}}
\left({\bf k}
\right)
\right]\!
{\tilde {\underline {\underline \Lambda}}}\left({\bf k}
\right)
\!  \cdot\! \left\{
{\tilde {\bf J}}',{\tilde {\bf M}}'
\!\right\}
\!\!  \left[
{\tilde {\bf F}}\left({\bf k}\right)\right]\!,\label{eq:source}\\
\left\{
{\tilde  {\underline {\underline \varepsilon}}}, {\tilde  {\underline {\underline \mu}}}
\right\}({\bf k})&\!\!\!\!=\!&
\!\!\det^{~~~~~-1}\!\left[{\tilde {\underline {\underline \Lambda}}}
\left({\bf k}
\right)
\right]\!
{\tilde {\underline {\underline \Lambda}}}\left({\bf k}
\right)
\cdot
{\tilde {\underline {\underline \Lambda}}}^T\left({\bf k}
\right),
\label{eq:tensors}
\end{eqnarray}
\label{eq:transf}
\end{subequations}
with $\mbox{det}(\cdot)$ denoting the determinant, and the superfix $^{-T}$ denoting the inverse transpose.

A few general considerations are in order. First, we note that the relationships in (\ref{eq:transf}) formally resemble those encountered in the standard (spatial-domain) TO approach \cite{Pendry}, and trivially reduce to them 
in the particular case of {\em linear} spectral mapping [i.e., ${\bf k}$-independent ${\tilde {\underline {\underline \Lambda}}}$ in (\ref{eq:mapping})], which is fully equivalent to the {\em local} coordinate mapping ${\bf r}'={\tilde {\underline {\underline \Lambda}}}^{-1}\cdot {\bf r}$. However, for a general {\em nonlinear} spectral mapping in (\ref{eq:mapping}), the resulting constitutive tensors in (\ref{eq:tensors}) are always ${\bf k}$-dependent, i.e., the associated constitutive relationships are {\em nonlocal}.
Similar to the spatial-domain TO approach, the spectral field/source transformations in (\ref{eq:field}) and (\ref{eq:source}) may be used to systematically design a desired response in a fictitious curved-coordinate spectral domain (in terms of a given nonlocal transformation of a reference field/source distribution in vacuum). Such response may be equivalently obtained in an actual physical space filled up by a nonlocal transformation medium whose constitutive ``blueprints'' are explicitly given by (\ref{eq:tensors}). Restrictions on the coordinate mapping in (\ref{eq:mapping}) may be imposed so as to enforce specific physical properties, such as passivity and/or reciprocity. For instance, it can readily be verified that the Hermitian condition ${\tilde {\underline {\underline \Lambda}}}^T\left({\bf k}\right)={\tilde {\underline {\underline \Lambda}}}^*\left({\bf k}\right)$ yields a lossless medium, whereas the center-symmetry condition ${\tilde {\underline {\underline \Lambda}}}\left({\bf k}\right)={\tilde {\underline {\underline \Lambda}}}\left(-{\bf k}\right)$ yields a reciprocal medium.

%
\begin{figure*}
\begin{center}
\includegraphics [width=14cm]{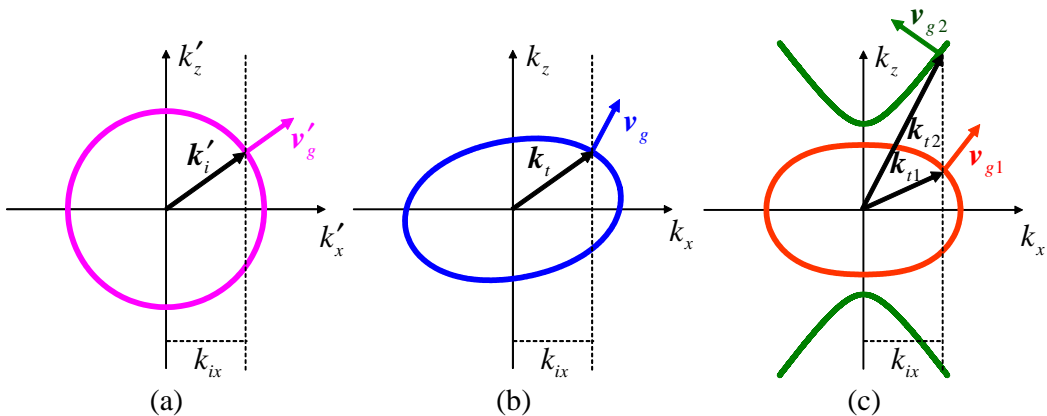}
\end{center}
\caption{(Color online) Schematic of EFCs pertaining to the auxiliary vacuum space (a) and the deformed versions obtained via a single-valued (b) and double-valued (c) spectral coordinate mapping. Also shown are the wavevectors and group velocities pertaining to a refraction scenario (details in the text).}
\label{Figure1}
\end{figure*}
In the spatial-domain TO, the choice of the coordinate transformation is guided by intuitive geometrical considerations essentially based on the geodesic path of light rays. Likewise, our nonlocal TO approach admits a geometrical interpretation in terms of direct manipulation of the dispersion characteristics via deformation of the {\em equi-frequency contours} (EFCs). While perhaps less intuitive, such interpretation is equally insightful and powerful, as 
the geometrical characteristics (e.g., asymptotes, symmetries, inflection points, single/multi-valuedness) of the EFCs fully determine the kinematical (wavevector and velocity) properties of the wave propagation and reflection/refraction \cite{Lock}.
Figure \ref{Figure1} schematically illustrates this interpretation with reference to an ($x,z$) two-dimensional (2-D) scenario where the EFC pertaining to the vacuum space is given by ${k'}_x^2+{k'}_z^2=k_0^2$ [Fig. \ref{Figure1}(a)], i.e., a circle of radius $k_0=\omega/c_0$ (the vacuum wavenumber, with $c_0=1/\sqrt{\varepsilon_0\mu_0}$ denoting the corresponding speed of light). Figures \ref{Figure1}(b) and \ref{Figure1}(c) show two qualitative examples of transformation-medium EFCs, 
\beq
{\tilde F}_x^2(k_x,k_z)+{\tilde F}_z^2(k_x,k_z)=k_0^2,
\label{eq:implicit}
\eeq 
which, depending on whether the mapping in (\ref{eq:mapping}) is {\em single}- or {\em double}-valued, may feature a moderate deformation [Fig. \ref{Figure1}(b)] or the appearance of an extra branch [Fig. \ref{Figure1}(c)], respectively.
Our geometrical interpretation therefore establishes a straightforward connection between the multi-valued character of the mapping and the presence of additional extraordinary waves.

Moreover, for the same 2-D scenario above, assuming a single-valued mapping and letting $\Phi_0(x)$ and $\Phi_d(x)$ the aperture distributions of a transverse (electric or magnetic) field at the input ($z=0$) and output ($z=d$) planes, respectively, in an unbounded material space,
the corresponding (1-D) spatial spectra will be related via
\beq
{\tilde \Phi}_d(k_x)=\exp\left(ik_zd\right){\tilde \Phi}_0(k_x)\equiv{\tilde T}(k_x){\tilde \Phi}_0(k_x),
\label{eq:PW}
\eeq
where the first equality arises from straightforward plane-wave algebra, and we have assumed
\beq
k_z=-\frac{i}{d}\log\left[{\tilde T}(k_x)\right].
\label{eq:explicit}
\eeq
Equation (\ref{eq:PW}) may be interpreted as an input-output relationship of a shift-invariant linear system, in terms of the {\em modulation transfer function} ${\tilde T}(k_x)$. Within the framework of our approach, Eq. (\ref{eq:explicit}) directly defines in {\em explicit} form the EFC shape that is needed in order to engineer a desired field-transformation effect between the input and output planes. Accordingly, the possibly simplest spectral mapping (\ref{eq:mapping}) from the auxiliary vacuum space that can yield [via (\ref{eq:implicit})] such desired shape is
\beq
{\tilde F}_x(k_x)\!=\!\sqrt{k_0^2+\frac{1}{d^2}\log^2\left[{\tilde T}(k_x)\right]},~~
{\tilde F}_z(k_z)\!=\!k_z.
\eeq 

The above examples highlight the intriguing perspectives of engineering the dispersion properties of the transformation medium via the vector mapping function ${\bf {\tilde F}}$ in (\ref{eq:mapping}).
Clearly, the practical applicability of the approach relies on the possibility to synthesize anisotropic, nonlocal artificial materials which, within given frequency and wavenumber ranges, suitably approximate the blueprints in (\ref{eq:tensors}). While, compared with the better established synthesis of anisotropic, spatially-inhomogeneous materials required by standard spatial-domain TO, this is significantly more challenging from the technological viewpoint, interesting nonlocal effects may still be engineered relying on simple artificial materials for which nonlocal homogenized models are available in the literature. 

\begin{figure*}
\begin{center}
\includegraphics [width=14cm]{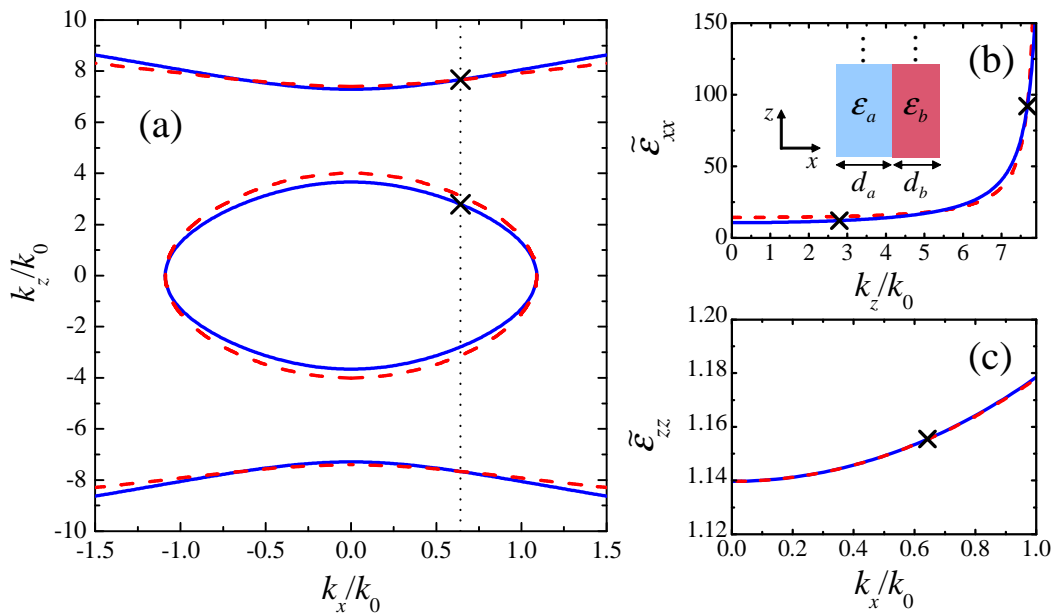}
\end{center}
\caption{(Color online) Examples of EFCs (a) and constitutive parameters (b,c) pertaining to a refraction scenario featuring the splitting of a plane wave with incidence angle $\theta_i=40^o$  into two transmitted waves with angles $\theta_{t1}=70^o$ and $\theta_{t2}=-45^o$. Blue (solid) curves represent the TO-based blueprints, obtained from (\ref{eq:Fmap}) and (\ref{eq:rational}) with $a_0=0.877$, $a_2=-0.0289 k_0^{-2}$, $b_0=0.0934$, $b_2=-0.0014 k_0^{-2}$. 
Red (dashed) curves pertain to the synthesized 1-D PC (unit-cell shown in the inset) with $\varepsilon_a=2.752$, $d_a=0.0668 \lambda_0$, $\varepsilon_b=-2.082$, $d_b=0.0332 \lambda_0$ (see \cite{auxiliary} for details). The vertical dotted line in (a) defines the incident wavenumber $k_{ix}$, from which the transmitted wavevectors (marked with crosses) are determined.}
\label{Figure2}
\end{figure*}

As an illustrative application example, we outline a step-wise procedure to design a nonlocal transformation-medium half-space so that a transversely-magnetic (TM) polarized (i.e., $y$-directed magnetic field) plane wave with assigned
wavevector ${\bf k}'={\bf k}_i$ [with angle $\theta_i$ from the $z$-axis and associated group velocity ${\bf v}'_g({\bf k}_i)=c_0{\bf k}_i/|{\bf k}_i|$, cf. Fig. \ref{Figure1}(a)] impinging from vacuum is split into {\em two} transmitted waves with prescribed directions, i.e., group velocity forming an angle $\theta_{t1}$ and $\theta_{t2}$, respectively, with the $z$-axis.
As schematically illustrated in Fig. \ref{Figure1}, the wavevector(s) ${\bf k}_t$ pertaining to the wave(s) transmitted into the transformation-medium half-space may be readily determined as the image(s) of the incident wavevector ${\bf k}_i$ in the deformed ECF(s) (\ref{eq:implicit}) subject to the tangential-wavevector continuity $k_{tx}=k_{ix}$ and to the radiation condition $\mbox{Re}(k_{tz})>0$. For a given transmitted wavevector ${\bf k}_t$, the corresponding group velocity (normal to the deformed EFC) is given by 
\beq
\left.{\bf v}_g({\bf k}_t)\equiv\frac{\partial \omega}{\partial{\bf k}}\right|_{{\bf k}_i}=\pm\frac{c_0{\tilde {\underline {\underline J}}}^T({\bf k}_t)\cdot{\tilde {\bf F}}\left({\bf k}_t\right)}{|{\tilde {\bf F}}\left({\bf k}_t\right)|},
\label{eq:vg}
\eeq
with ${\tilde {\underline {\underline J}}}({\bf k})\equiv \partial{\bf k}'/\partial{\bf k}$ denoting the Jacobian matrix of the transformation in (\ref{eq:mapping}), and the $\pm$ sign dictated by the radiation condition $\mbox{Re}(v_{gz})>0$.
We first need to determine a {\em double-valued} spectral-domain transformation (\ref{eq:mapping}) capable of mapping [via (\ref{eq:implicit})] the incident wavevector ${\bf k}_i$ into two transmitted wavevectors ${\bf k}_{t1}$ and ${\bf k}_{t2}$ characterized by a conserved tangential (i.e., $x$-) component and the desired group-velocity directions, i.e.,
\begin{subequations}
\begin{eqnarray}
k_{t1x}=k_{t2x}&=&k_{ix}=k_0\sin\theta_i,
\label{eq:conservation}\\
\displaystyle{\frac{v_{gx}({\bf k}_{t1,2})}{v_{gz}({\bf k}_{t1,2})}}&=&\tan\theta_{t1,2},
\label{eq:vg1}
\end{eqnarray}
\end{subequations}
with ${\bf v}_g$ given by (\ref{eq:vg}). A simple analytical solution to this functional problem may be obtained by assuming a variable-separated algebraic mapping of the form 
\beq
{\tilde F}_x(k_x)\!=\!k_x\sqrt{a_0+a_2 k_x^2},~~{\tilde F}_z(k_z)\!=\!k_z\sqrt{b_0+b_2 k_z^2},
\label{eq:Fmap}
\eeq 
with the coefficients $a_0, a_2, b_0$ and $b_2$ to be determined. 
First, by substituting (\ref{eq:Fmap}) in (\ref{eq:implicit}) [and taking into account (\ref{eq:conservation})], the above choice allows analytical closed-form calculation of the transmitted wavevectors ${\bf k}_{t1}$ and ${\bf k}_{t2}$, via straightforward solution of a biquadratic equation (see \cite{auxiliary} for details). Next, by substituting ${\bf k}_{t1}$ and ${\bf k}_{t2}$ in (\ref{eq:vg1}) [with (\ref{eq:vg}) and (\ref{eq:Fmap})], we obtain an analytically-solvable system of two algebraic equations, whose solutions constrain two coefficients (say $b_0$ and $b_2$) in (\ref{eq:Fmap}), thereby defining a family of (infinite) coordinate transformations which, for the given incidence conditions, yield the prescribed kinematical characteristics ($\theta_{t1}$ and $\theta_{t2}$) of the two transmitted waves (see \cite{auxiliary} for details). For instance, assuming an incidence angle $\theta_i=40^o$, and two transmission angles $\theta_{t1}=70^o$ (i.e., positive refraction) and $\theta_{t2}=-45^o$ (i.e., negative refraction), Fig. \ref{Figure2}(a) shows [blue (solid) curves] the double-valued EFCs pertaining to one such transformation (with parameters given in the caption). 
%

Note that the two seemingly free parameters in the transformation ($a_0$ and $a_2$) may be in principle exploited to enforce additional conditions (e.g., at a different frequency). Nevertheless, we found it useful to maintain the flexibility endowed by such degrees of freedom in order to facilitate the engineering of the transformation medium required. Within this framework,
from (\ref{eq:tensors}), we first need to define the tensor operator ${\tilde {\underline {\underline \Lambda}}}$ associated [via (\ref{eq:mapping})] to the vector mapping ${\tilde {\bf F}}$ in (\ref{eq:Fmap}). The possibly simplest choice \cite{endnote2} is the {\em diagonal} form ${\tilde {\underline {\underline \Lambda}}}=\mbox{diag}[{\tilde F}_x/k_x,{\tilde \Lambda}_{yy},{\tilde F}_z/k_z]$, where, for the assumed TM-polarization, the component ${\tilde \Lambda}_{yy}$ represents a degree of freedom which may be judiciously exploited so as to simplify the structure of the arising transformation medium. Paralleling the spatial-TO approach, a desirable property, which may strongly facilitate the scalability towards optical frequencies, is an {\em effectively non-magnetic} material (i.e., ${\tilde \mu}_{yy}=1$) \cite{Cai}. This is readily achieved by choosing ${\tilde \Lambda}_{yy}={\tilde F}_x{\tilde F}_z/(k_xk_z)$, which yields [from (\ref{eq:tensors}) and (\ref{eq:Fmap})] a uniaxial anisotropic medium whose relevant permittivity components assume a particularly simple variable-separated {\em rational} form
\beq
{\tilde \varepsilon_{xx}}(k_z)=\frac{1}{b_0+b_2 k_z^2},~~
{\tilde \varepsilon_{zz}}(k_x)=\frac{1}{a_0+a_2 k_x^2},
\label{eq:rational}
\eeq
whose behavior is shown [blue (solid) curves] in Figs. \ref{Figure2}(b) and \ref{Figure2}(c) for the same parameters as above.

Interestingly, the parameterization in (\ref{eq:rational}) closely resembles the nonlocal homogenized constitutive relationships derived in \cite{Elser} for a 1-D multi-layered photonic crystal (PC), thereby suggesting that our TO-based blueprints may be approximated, at a given frequency and within limited spectral-wavenumber ranges, by such a simple artificial material. 
Accordingly, we consider a 1-D PC made of alternating layers (periodically stacked along the $x$-axis) of homogeneous, isotropic materials, with relative permittivity $\varepsilon_a$ and $\varepsilon_b$, and thickness $d_a$ and $d_b$, respectively [see the inset in Fig. \ref{Figure2}(b)].
In \cite{Elser}, the parameters of the homogenized uniaxial medium [cf. (\ref{eq:rational})] were determined by matching its dispersion law with the McLaurin expansion of the exact Bloch-type dispersion law of the PC up to the fourth order in the arguments.
Such nonlocal homogenized model establishes a simple analytical connection with the PC parameters, and is therefore very useful in our synthesis procedure. However, in our case we developed a modified version [similar to (\ref{eq:rational}), but higher-order in $k_z$], in order to accurately capture the parameter variations over the dynamical ranges involved (see \cite{auxiliary} for details). 

As a last step, we developed a semi-analytical procedure for determining the parameters of the PC approximant, based on the matching between the above nonlocal homogenized model and the TO-based blueprints in (\ref{eq:rational}) (see \cite{auxiliary} for details). For the example considered, the above synthesis yields a PC with $\varepsilon_a=2.752$, $d_a=0.0668 \lambda_0$, $\varepsilon_b=-2.082$, $d_b=0.0332 \lambda_0$ (with $\lambda_0=2\pi/k_0$ denoting the vacuum wavelength).  
Figure \ref{Figure2} compares the corresponding exact (i.e., Bloch) EFCs and (nonlocal homogenized) constitutive parameters [red (dashed) curves] with the TO-based blueprints. A satisfactory agreement is observed, especially within neighborhoods of the prescribed transmitted wavevectors ${\bf k}_{t1}$ and ${\bf k}_{t2}$ (marked with crosses), which are directly relevant to the refraction scenario of interest. 
%
\begin{figure*}
\begin{center}
\includegraphics [width=14cm]{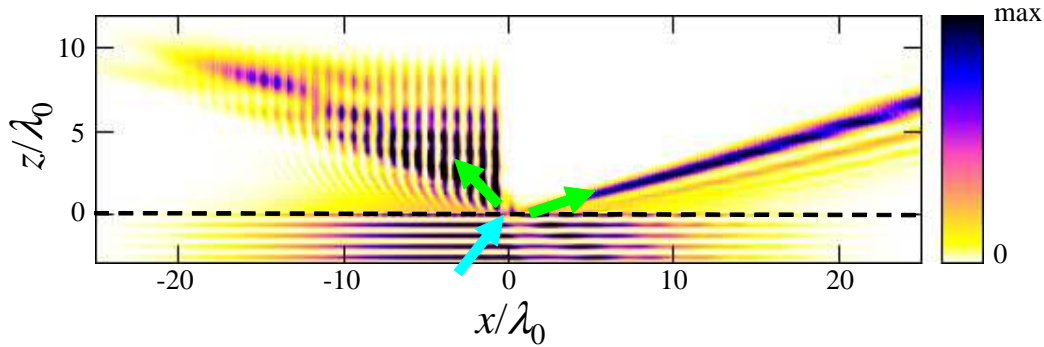}
\end{center}
\caption{(Color online) FDTD-computed magnetic-field intensity ($|H_y|^2$) map illustrating the transmission of a collimated Gaussian beam in a finite-size PC (with parameters as in Fig. \ref{Figure2} and details in \cite{auxiliary}). The dashed line indicates the vacuum-PC interface, whereas the cyan and green thick arrows indicate the incident-beam and (prescribed) transmitted-beam directions, respectively.}
\label{Figure3}
\end{figure*}

As an independent validation of the above synthesis procedure, we carried out a finite-difference-time-domain (FDTD) simulation \cite{Taflove} involving a finite-size PC slab (see \cite{auxiliary} for details). Figure \ref{Figure3} shows a field map illustrating the splitting of an incident wide-waisted Gaussian beam into two transmitted beams with directions consistent with the prescribed ones. It is worth pointing out that the positively-refracted beam arises from {\em local} effects, and may be predicted by standard effective-medium modeling. Conversely, as clearly visible in Fig. \ref{Figure3}, the negatively-refracted beam originates from the excitation and coupling of surface-plasmon-polaritons propagating along the interfaces between the negative-permittivity and dielectric layers of the PC, and therefore constitutes an additional extraordinary wave which can only be predicted by {\em nonlocal} modeling.

For this particular example, one may argue that a semi-heuristic identification of the required artificial material and a direct optimization of its structural parameters (so as to approximate the desired EFCs) may have been as effective as the design based on the TO theory developed here. However, in a more general application scenario, for which a more complicated dispersion relation and a larger number of structural parameters may be desired, a direct optimization approach would require iterative numerical full-wave solutions of source-free EM problems, and may therefore become computationally unaffordable. Conversely, the inverse process proposed here, while seemingly more involved, does not require full-wave modeling, and may still be carried out in a computationally-mild semi-analytical fashion as a parameter matching between the TO-based blueprints and the nonlocal homogenized model, similar to the above example.
 
In conclusion, we have introduced and validated a spectral-domain-based TO framework which admits a physically-incisive and powerful geometrical interpretation in terms of EFC deformation. Our approach crucially relies on the availability of nonlocal homogenized models, and allows the systematic synthesis of spatially-dispersive transformation media capable of yielding prescribed nonlocal field-transformation effects. Given the current research trend in metamaterial homogenization, with a variety of rigorous theories that allow the description of the wave interaction in metamaterials in the spectral domain \cite{Silveirinha1,Silveirinha2}, and a correspondingly growing ``library'' of nonlocal homogenized models, we expect that our approach may rapidly become an exciting option for spatial dispersion engineering in the near future.
Also of great interest is the exploration of {\em nonreciprocal} effects, which our approach can naturally handle via the use of non center-symmetric coordinate transformations.

\newpage


\begin{thebibliography}{99}
\bibitem{endnote2}{Note that such correspondence is not unique, reflecting the fact that different media may yield the same dispersion properties.}
\end{thebibliography}
\end{document}